# Cavitation bubble dynamics and sonochemiluminescence activity inside sonicated submerged flow tubes

Busra Ekim Sarac, Dwayne Savio Stephens, Julian Eisener[1], Juan Manuel Rosselló, Robert Mettin[*]

*Drittes Physikalisches Institut, Georg-August-Universität Göttingen, Friedrich-Hund-Platz 1, 37077 Göttingen, Germany*

**Abstract**

Bubble dynamics and luminol emissions of cavitation in sub-millimeter-sized PFA flow tubes, submerged in an ultrasonic bath reactor, are studied at 27 kHz driving frequency. Nucleation of cavitation inside the tubes only takes place via a free interface, realized here in form of an alternating water-air slug flow. High-speed recordings show that cavitation bubbles in the water slugs often develop localized structures in form of clusters or bubble "plugs", and that such structures can be seeded via a single pinch-off from the free interface. Within the structures, bubbles strongly interact and frequently undergo merging or splitting events. Due to the mutual interaction and resulting motion, bubbles often collapse with a fast displacement, suggesting jetting dynamics. Bubble compression ratios are estimated on basis of observed individual bubble dynamics and numerical fitting by a single bubble model. The resulting peak temperatures around 3000 K allow for dissociation of water vapor. This is in accordance with observed sonochemiluminescence from luminol, originating from active bubble zones in the tubes.

1 present address: Institut für Physik, Otto-von-Guericke-Universität Magdeburg, Universitätsplatz 2, 39106 Magdeburg, Germany

* corresponding author, email address: robert.mettin@phys.uni-goettingen.de (R. Mettin)

**Highlights**

- High-speed observation of cavitation bubble structures inside sub-millimeter PFA tubes
- Bubble nucleation only in water/air slug flow and via entrained gas from the free interface
- Active bubbles mainly form localized clusters or plugs with the order of 30 to 200 strongly interacting bubbles
- Sonochemiluminescence recordings and numerical fits of observed bubble dynamics allow for assessment of chemical activity

## 1. Introduction

Process intensification (PI), constitutively defined as a continuous processing via flow reactors, is the most promising innovative development in fine-chemical and pharmaceutical industries in the last decades [1]. The overall aim of PI is to improve product quality, control the process precisely, reduce waste, ease the scale up, reduce energy consumption and use the raw material efficiently [2] One of the recent developments in PI is the use of flow microreactors [3] Microreactors offer low characteristic length scale, high surface-volume ratio and increase in mixing due to internal circulation; thus, they have the advantage of enhancements of heat and mass transfer coefficients and increase in energy conversion efficiency. Moreover, small volumes can be cost efficient and environment friendly due to the reduction in the size of the equipment, less energy consumption, easier and safer handling of hazardous chemicals, and better controlling ability of reactions taking place at high temperatures and pressures [4]

Innovative research has been focusing on combining microreactors with non-classical, non-contact and sustainable energy sources [5][6]. These include for instance electrostatic fields, microwaves, plasma radiation, and ultrasound. Here we focus on the latter method: irradiation of flow reactors with high frequency acoustic waves, i.e. ultrasound [7].

The general case of chemistry initiated and/or enhanced by ultrasound is termed *sonochemistry*, which is also considered a green and sustainable chemistry [8][9].

Usually the energy of ultrasound interacts with the molecules in a liquid through acoustic cavitation [10][11]. The strong ultrasonic driving pressure field forms and expands cavitation bubbles in the tension (negative pressure) phase, typically at interfaces or floating weak spots called nuclei. In the subsequent overpressure phase of the field, the bubbles implode and can develop high pressures and temperatures inside of the order of T≈5000 K and P≈1000 bar. The rapid heating can trigger chemical reactions in the gas phase [8][12][13], but also in the liquid phase if liquid enters the collapsing bubble [14] [15].

The chemical effect of cavitation in aqueous environments is often linked to the radical formation in water vapor. Hydrogen atoms and free hydroxyl radicals ($OH^{\bullet}$) are formed as a result of high temperatures and pressures inside the bubble during the last stage of its collapse. The hydroxyl radicals can be detected by visible blue light emission from luminol, termed *sonochemiluminescence (SCL)* [16][17][18]. SCL has to be distinguished from native *sonoluminescence (SL)* that is generated in the "hot spot" of the imploding bubble by thermal and plasma radiation [19] [20].

Ultrasound frequencies in sonochemistry range from 20 kHz to several MHz, and amplitudes and operation modes (e.g. pulsed vs. continuous) can be varied. In the framework of microreactors, numerous designs for sonochemistry have been reported in the last decade for different chemical reactions [17] [21] [22] Among them are liquid-liquid extraction [21] and gas-liquid mass transfer intensification [23] by a direct contact method of flow tube and Langevin transducer, OH radical formation in a polydimethylsiloxane-based microfluidic reactor in contact with the driving piezo ceramic [17] handling solid forming reactions by a teflon stack microreactor with an integrated piezoelectric actuator [24] or in a microchannel on top of a Langevin transducer [25] crystallization through sonication of a flow cell by an integrated piezo ceramic [26], fabrication of nanoparticle-coated microbubbles through microfluidic channels irradiated by an ultrasonic horn [27] and crystallization of acetylsalicylic acid through milichannels sonicated as well by a horn transducer [28].

As well as various methods to assemble the microreactor exist, different compositions and dimensions of channels, employed according to the reactants and the nature of chemical reaction, have been recently reported. These include glass channels attached to a microscope slide [17], PE/5 channels utilized with an ultrasonic horn [27], PDMS channels via lithography [29], silicon channels via micromachining [18] or PTFE channels in a teflon microreactor [24].

Possibly the most simple way of sonication is the submersion of sound transmissible flow tubes in a larger batch reactor or cleaning bath [30], and we are reporting here on such a setup. Advantages include easy installation, potentially large irradiated volumes and long residence times (i.e. long tube lengths affected by the irradiation), and the possibility of temperature control via the coupling liquid in the bath. Furthermore, the setup can be fully transparent, which is utilized here for direct imaging of cavitation bubbles and for assessment of SCL from luminol. Drawbacks may occur due to larger ("macroscopic") installation equipment (the bath), limited frequencies in the lower ultrasonic range, lower energy efficiency, or less controlled and unstable operation. The latter point arises since several parameters and conditions of bath and outer (coupling) liquid can affect the energy transfer to the working liquid in the tube, for instance exact fixing position of the tube, filling height, temperature and dissolved gas content of the outer liquid, or dissipation by cavitation therein. Thus these conditions might need additional control for reliable operation. Still, the submerged tube configuration is a prototypical one that is oftentimes used and that can give general information on properties of cavitation in irradiated tubes or channels. Moreover, modifications and variants of the "simple" submerged tube can alleviate some of the listed downsides in customized configurations, e.g. with smaller coupling liquid volumes or higher frequency transducers.

In the present study, a perfluoroalkoxy alkane (PFA) tube is submerged into a water-filled ultrasonic bath. PFA is a hydrophobic polymer offering high thermal and chemical resistance. It has also reasonably high flexibility and low bending radius which allow ease of reactor construction. Additionally, PFA provides an opportunity of observing the events taking place inside the channel due to its high transparency. Finally, the acoustic impedance of PFA is close to that of water [31], allowing for nearly transparent sound propagation through the tube walls into the reactive liquid volume.

Our setup and experimental procedures are described in more detail in Section 2. Some theoretical and numerical aspects, used for the evaluation of observations, are discussed in Section 3. The main results are presented in Section 4: luminol emission measurements, high-speed videography of cavitation bubble structures in the tubes, estimates of bubble collapse compression ratios on basis of numerical fits by a single bubble model, and observation of nucleation events via a free gas/liquid interface. A conclusion is given in Section 5.

## 2. Experimental part

A schematic drawing of the setup is shown in Figure 1. We employ an in-house made rectangular transparent bath reactor with transparent polycarbonate walls (makrolon, thickness of 6 mm) and open on top. It is sonicated at 27200 Hz by a piezoceramic Langevin transducer (Elmasonic, Germany) glued to a steel plate that is forming the bottom wall. Dimensions of the bath are 140×50×150 mm$^3$, (l× w ×h), and filtered, non-degassed water at room temperature (20°C) as the coupling liquid is filled up to 60 mm height. The transducer is driven by a frequency generator (Tektronix AFG 3021) via a power amplifier (E&I Ltd., 1040L, USA) and an in-house built impedance matching box. Cavitation structures in this device under various conditions have been described before in [32]. Here we submerge a perfluoroalkoxy alkane (PFA) tube (BOLA, Germany) of inner diameter $d_i = 1/32$'' ($\approx 0.8$ mm) and outer diameter $d_o = 1/16$'' ($\approx 1.6$ mm) from top into the water, whereby the tube undergoes three loops of approximately 50 mm diameter each. The tube is fixed centrally over the transducer by a holder, and it is connected to a pair of syringe pumps (ProSense, Multi-Phaser$^{TM}$ NE-500, The Netherlands) on the inlet side. The pumps can alternatively supply air or aqueous luminol solution via a T-junction into the tube, the outlet side is open. By switching the pumps, alternating slugs of air and luminol solution of about 1 cm length each are produced inside the submerged tube. During ultrasound operation, the slug lengths and air gaps can change due to mass exchange by droplet ejection and air entrainment, both described below. For the luminol and bubble measurements, the flow is stopped, i.e. the gas/liquid slug distribution in the tube is stationary. SCL measurements are carried out with 0,1 mM luminol solution. Luminol, i.e. 3-aminophthalhydrazide (98%, Fluka, USA) is dissolved in NaOH solution (32 wt.%, Atotech, Germany) and deionized water at room temperature, and pH is adjusted to 11.0. Luminol light emissions are observed by a digital SLR camera (Nikon D700, Japan) in dark room conditions under long exposure (30 seconds). Cavitation bubbles are visualized by a high-speed camera (Photron , Fastcam SA5, Japan) via a long-distance microscope (K2/SC, Infinity, USA ). Illumination is provided by an intense cw white light source (Sumita LS-352A, Japan). Bubbles appear dark in front of a bright background.

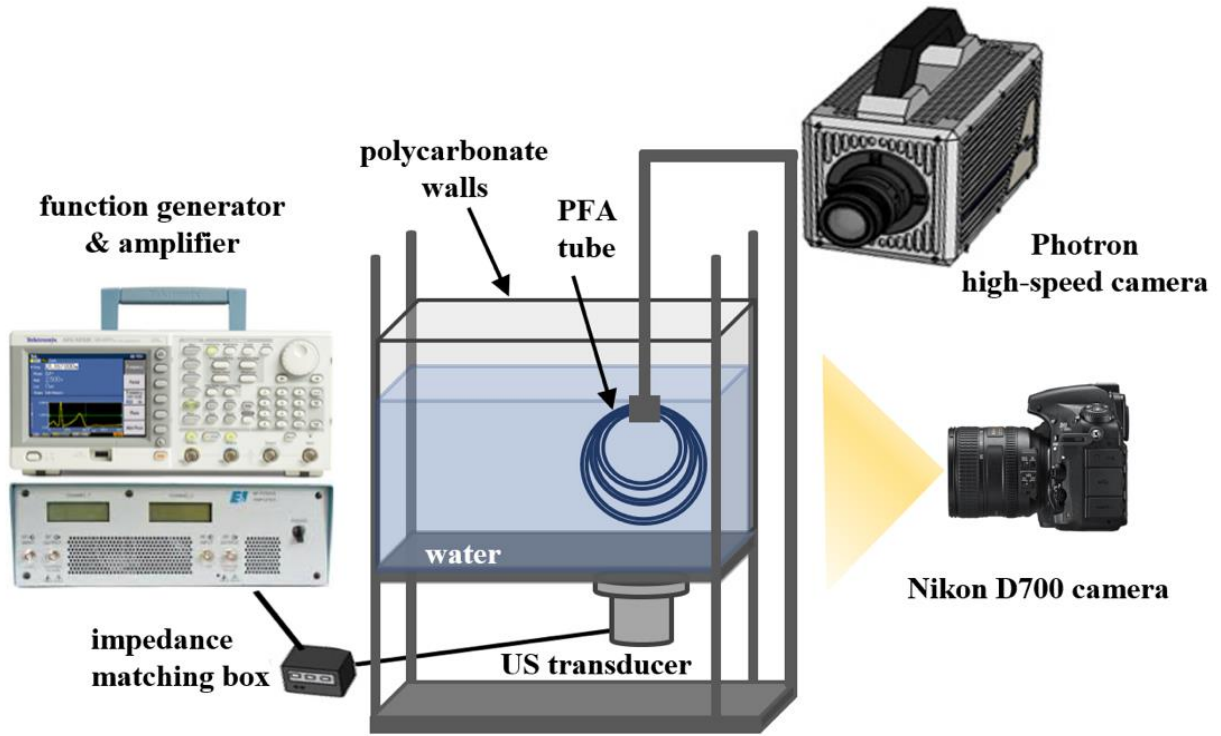

*Figure 1: Schematic drawing of the setup. Three loops of PFA tubing filled with aqueous luminol solution (0.1mM) are submerged in an ultrasonic bath driven at 27.2 kHz and observed by a digital camera. Alternatively, the interior of the tubes is visualized via a long-distance microscope by a high-speed camera with respective illumination.*

### 3. Theory and simulation

The ultrasonic field in the resonator develops a standing wave pattern and causes cavitation in the outer (coupling) liquid. We model the field by a modified Helmholtz equation, according to the model by Louisnard [33][34] that takes nonlinear dissipation by homogeneously distributed bubbles into account. In this approach, the bubbles just form passive nuclei below a certain acoustic pressure amplitude threshold, but evolve into strongly dissipating cavitation bubbles beyond that. Parameters are standard values for water and a bubble void fraction of $\beta = 5 \cdot 10^{-6}$ with monodisperse nuclei of equilibrium size $R_0 = 3 \mu m$. The model is solved with the finite element software Comsol (Comsol AB, Stockholm, Sweden) in a 3D domain with impedance boundary conditions at the container walls, and without the tube present. The pressure distribution for a transmitted power of 65 W is shown for a central plane in the resonator in Figure 2a. The figure also shows an overlaid image of the approximate tube's position. Due to bubble shielding in front of the transducer (where higher pressures occur), the calculation results in a standing wave field with moderate pressure amplitudes of maxima around $\hat{p}_a = 140$ kPa. Figure 2b, c gives the pressure values along a vertical line centrally above the transducer face, and a horizontal line parallel to the bottom at a height of 5 mm where the high-speed recordings took place. According to the field simulation, parts of the tube are crossing pressure antinodal regions, while others are passing through low pressure (nodal) zones. Therefore it is expected that those parts of the tube should not be emitting SCL which are near the nodes. This supposes, of course, that disturbances of the field by the tube walls and the air slugs are negligible.

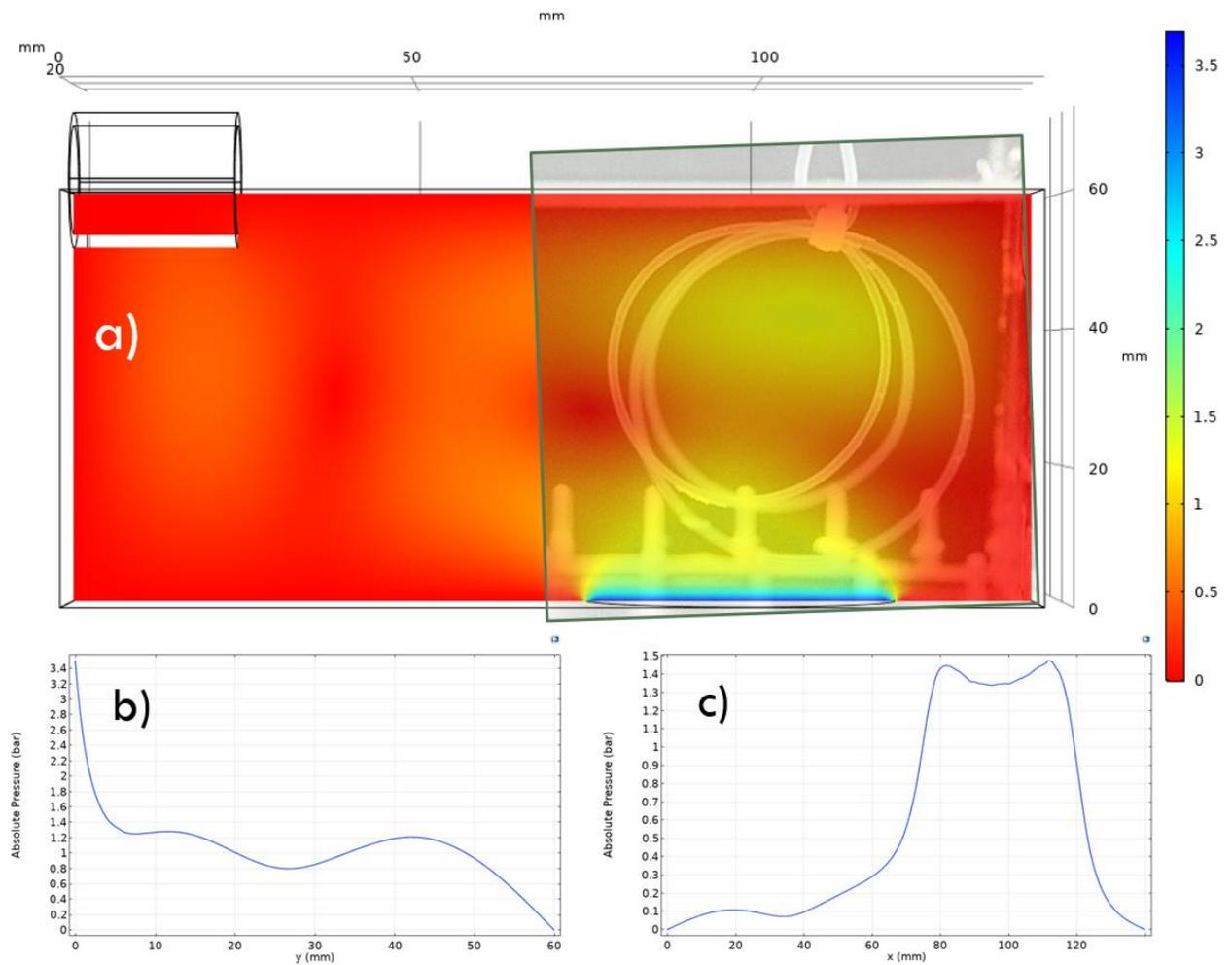

*Figure 2: (a) Simulated sound field in the reactor (color coded absolute pressure amplitude on a central vertical plane section, homogeneous void fraction $\beta = 5 \cdot 10^{-6}$, radiated power 65 W). An image of the tube loops is overlaid for illustration. (b) Calculated pressure amplitude vertically above the transducer. (c) Calculated pressure amplitude horizontally at a height of 5 mm.*

While cavitation bubbles in the outer fluid are free to move when driven by primary Bjerknes forces [11][35] , bubbles inside the tube cannot cross the walls and are thus trapped. However, pressure gradients inside the tubes, both in longitudinal and in transverse direction, can push the bubbles along the tube or towards the walls. Furthermore, secondary Bjerknes forces that are acting on shorter distances [11][35] will work similarly as in the bulk liquid, with an additional potential attractive mirror-bubble effect near the tube walls [36] .Both transverse primary Bjerknes forces and secondary mirror-bubble Bjerknes forces can lead to preferential bubble locations near the tube walls.

When individual bubble oscillations are resolved sufficiently in the experiment, bubble collapse conditions can be obtained from simulated radius-time dynamics. Employing a numerical bubble model, the bubble equilibrium radius and the local pressure amplitude are fitted to reproduce observed data, as exemplified for trapped stationary bubbles [37]. If the system is unsteady as in typical multibubble systems, the method can still be useful for an estimate. Here we apply a backfolding method of a few observed bubble oscillation periods to a single acoustic period to improve the temporal resolution of the image recordings [38][39]. The dynamics is then fitted by a single bubble model, the Keller-Miksis model [40]:

$$R\ddot{R}(1 - M) + \frac{3}{2}\dot{R}^2\left(1 - \frac{M}{3}\right) + \frac{4\mu}{\rho}\frac{\dot{R}}{R} + \frac{2\sigma}{\rho R} = \frac{1+M}{\rho}p_l + \frac{R}{\rho c}\frac{dp_l}{dt} \quad (1)$$

$$p_l = \left(p_0 + \frac{2\sigma}{R_0}\right)\left(\frac{R_0}{R}\right)^{3\gamma} - \frac{2\sigma}{R} - \frac{4\mu\dot{R}}{R} - p_0 - p_{\text{ac}}(t) \quad (2)$$

The model describes the temporal evolution of the bubble radius $R(t)$ of a bubble with equilibrium radius $R_0$ under driving with $p_{ac}(t) = p_a\sin(2\pi f t)$. Here $p_a$ is the pressure amplitude (set to $\hat{p}_a = 140$ kPa), and $f$ is the acoustic frequency (27200 Hz). The equation includes compressibility effects by the Mach number of the bubble wall $M = \dot{R}/c$ , $c$ being the sound velocity in the liquid (1482 m/s). Further parameters are static pressure $p_0$ (100 kPa), surface tension $\sigma$ (0.0725 N/m), dynamic viscosity $\mu$ (0.001 Ns/m$^3$), and density of the liquid $\rho$ (1000 kg/m$^3$). The polytropic exponent is set to adiabatic gas compression, i.e. $\gamma = 1.4$ for air.

As measures from the simulations, maximum ($R_{max}$) and minimum radii ($R_{min}$) are identified, and compression ratios $R_0/R_{min}$ are determined under an adiabatic gas compression law. This results in an upper bound for the temperatures during bubble collapse, since heat conduction, molecule dissociation and ionization, and other energy sinks are neglected. We use the values as an estimate of the bubble collapse conditions, being aware that more extended and complex models will somehow lower the figures.

## 4. Results

### 4.1 Luminol emissions

For sonicating the tube being fully filled with liquid, we do not obtain any luminol signal, and we do neither observe bubbles in the tubes. This is in accordance with other reports on cavitation in small channels (e.g. Tandiono et al. for PDMA channels of 20 µm height and sonicated directly via the substrate at 100 kHz [17] Here it shows that also in the larger tubes no nucleation of cavitation bubbles occurs under completely filled conditions. Both luminol emission and visual bubbles, however, do appear for alternate filling with liquid and air. For imaging, the flow has been stopped to improve contrast since the SCL emissions were generally quite low. Then, the non-moving slugs of aqueous luminol solution partly show weak emission of the characteristic blue light, indicating production of OH radicals by cavitation [16]. The images shown in Figure 3a, b have been captured during a 30 s exposure. It can clearly be seen that SCL emissions only occur from the liquid inside the tube (Figure 3c, d). Since the slug lengths are somehow changing during sonication, there occur also emitting regions or gaps longer or shorter than about 1 cm. A closer inspection shows that not all liquid filled tube regions appear blue, i.e., weaker or no SCL at all might occur. Reasons can be an unfavorable position in the standing wave, and/or missing nucleation in the specific slug. Furthermore, some regions show stronger localized signals (whiter spots). This is in accordance to the localized bubble structures inside the slugs, as described below. We do not, though, observe a clear correlation of luminol emission with the high amplitude regions from the acoustic field simulation (Figure 3e, f). One does see a high signal at the top in Figure 3a, b and at the left bottom of Figure 3b, both corresponding to the calculated antinode zones. The bottom part in Figure 3a, however, is not emitting prominently, whereas parts of the tube positioned at middle height (where lower acoustic pressures are predicted) are partly emitting. Thus the field simulation might not be very accurate. Potentially, the presence of gas slugs is perturbing the field significantly due to reflections and has to be taken into account. This could also explain differences of emissions between Figure 3a and b. Interestingly, the cavitation phenomena generally show variations depending on liquid slug length and slug spacing (i.e., gas pocket length). This variability has also been observed before [41], and it is subject of future studies.

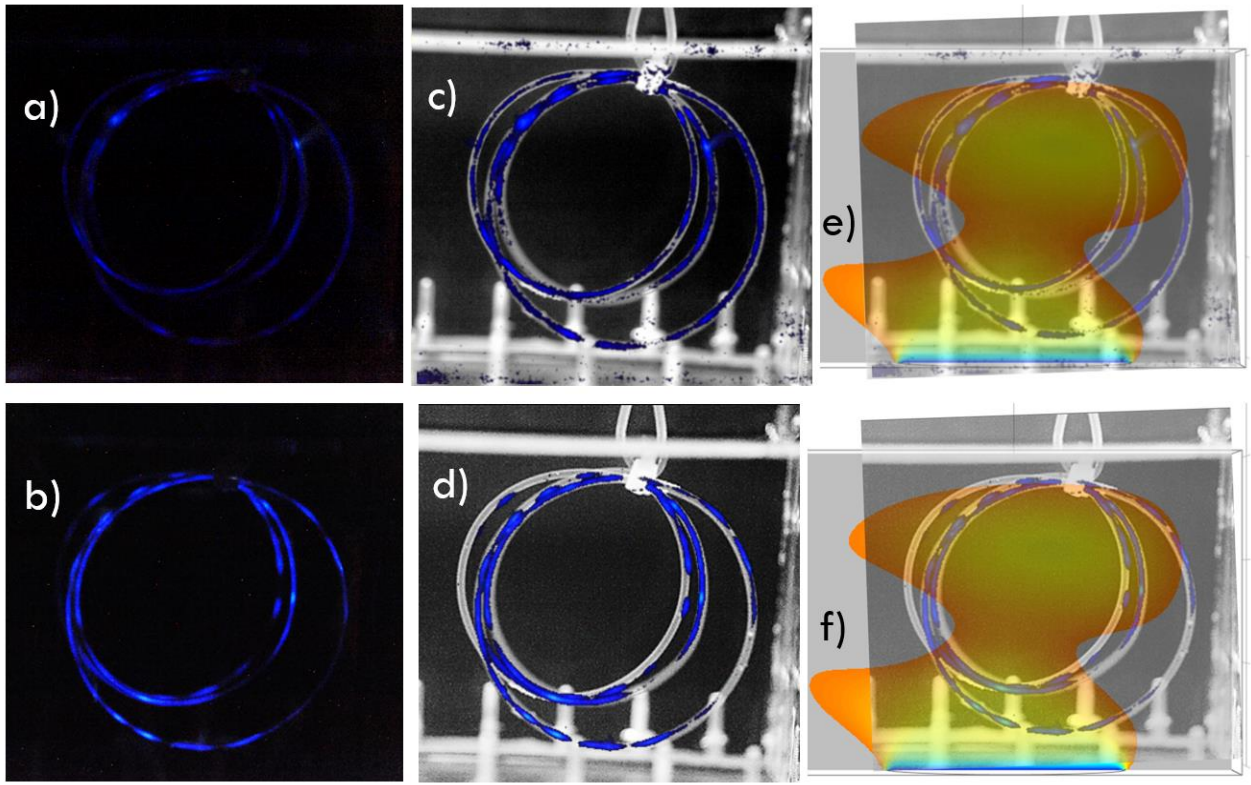

*Figure 3:* *Imaged luminol emissions and their relation to the submerged channels and the calculated sound field: Dark room exposures (30 s) for two different air/water slug positions are shown in a) and b). Overlays of contrast enhanced SCL images over the tube are given in c) and d); the overlays are partly not fitting perfectly since the tube is slightly shaking during operation. The approximate positions of the emitting regions and the tube in relation to the calculated standing wave pattern are shown in e) and f). The pressure amplitude is cut at 60 kPa, i.e., the orange to yellow regions represent amplitudes of 60 to 140 kPa.*

### 4.2 Bubble structures

Recordings of cavitation bubbles have been done in a section of the tube approximately 5 mm above the transducer (the lowest central tube part visible in Figure 3). The bubbles appear quite intermittent and in a certain variety, i.e., cavitation activity is far from homogeneous in time and space. Figure 4 illustrates prototypical bubble ensembles, all observed within a total recording time of about one second. In particular, we many times see wall attached clusters in form of a roughly half spherical aggregate at the bottom or top of the tube (Figure 4a, b). Other frequent structures are "plugs" of bubbles, i.e., a roughly rectangular region extending from top to bottom of the tube and with rather sharp limits at the sides (Figure 4c, d and e, f). It remains unclear if the plugs are similar to the wall attached clusters, but just seen from below or from top (i.e., attached to the front or back wall of the tube). Both structure types show a pronounced confinement of bubbles in longitudinal tube direction. The boundaries to the neighbored, nearly bubble free regions occur somehow sharper for plugs, which might hint to a structure actually different from a wall cluster. Less confined structures appear as well, and they form streamers that cross larger longitudinal sectors (Figure 4g, h), or fully dispersed bubble fields (Figure 4i, k). Inside the confined structures, cavitation bubbles interact strongly: merging or splitting take place every few acoustic cycles, sometimes during each oscillation period. Accordingly, the bubbles move, and frequently a collapse "jump" appears, at times with a resolved jetting event. In Figure 4l, m displacement and jetting can be discerned (marked by arrows; due to the long exposure time the bubble silhouette over half a period is visible as a grey shade). The rapid displacement and the merging events prevent oftentimes a clear re-identification of an individual bubble after collapse, even more since many bubbles disappear from the image during the collapse phase due to limited spatial resolution (about 5 micron/pixel). Within the more dispersed structures, bubbles show less frequent interaction, as expected from the larger inter-bubble distances. Still, collision events take place frequently, i.e., every few acoustic cycles. The numbers of identifiable and resolved bubbles in the structures range from about 30 to 200, but it has to be noted that the amount of visible bubbles within one structure is variable during the oscillation period. Few bubbles can be seen during the collapse phase (for limited resolution), and the highest number of bubbles occurs somehow between minimum and maximum expansion. At the fully expanded state, again the bubble number is decreased, partly only apparently due to optical overlap and shielding, partly due to true merging (compare also Fernandez Rivas et al. [42] for this phenomenon). To demonstrate the variability of bubble sizes and numbers, all structures in Figure 4 are shown in a nearly collapsed phase (first frame) and in the subsequent expansion phase (second frame). The void fraction in the collapsed cluster of Figure 4e is roughly estimated to about $2.5 \cdot 10^{-4}$, and it increases about 100-fold to $2.5 \cdot 10^{-2}$ in the expansion phase (Figure 4f). These numbers appears typical for the localized structures.

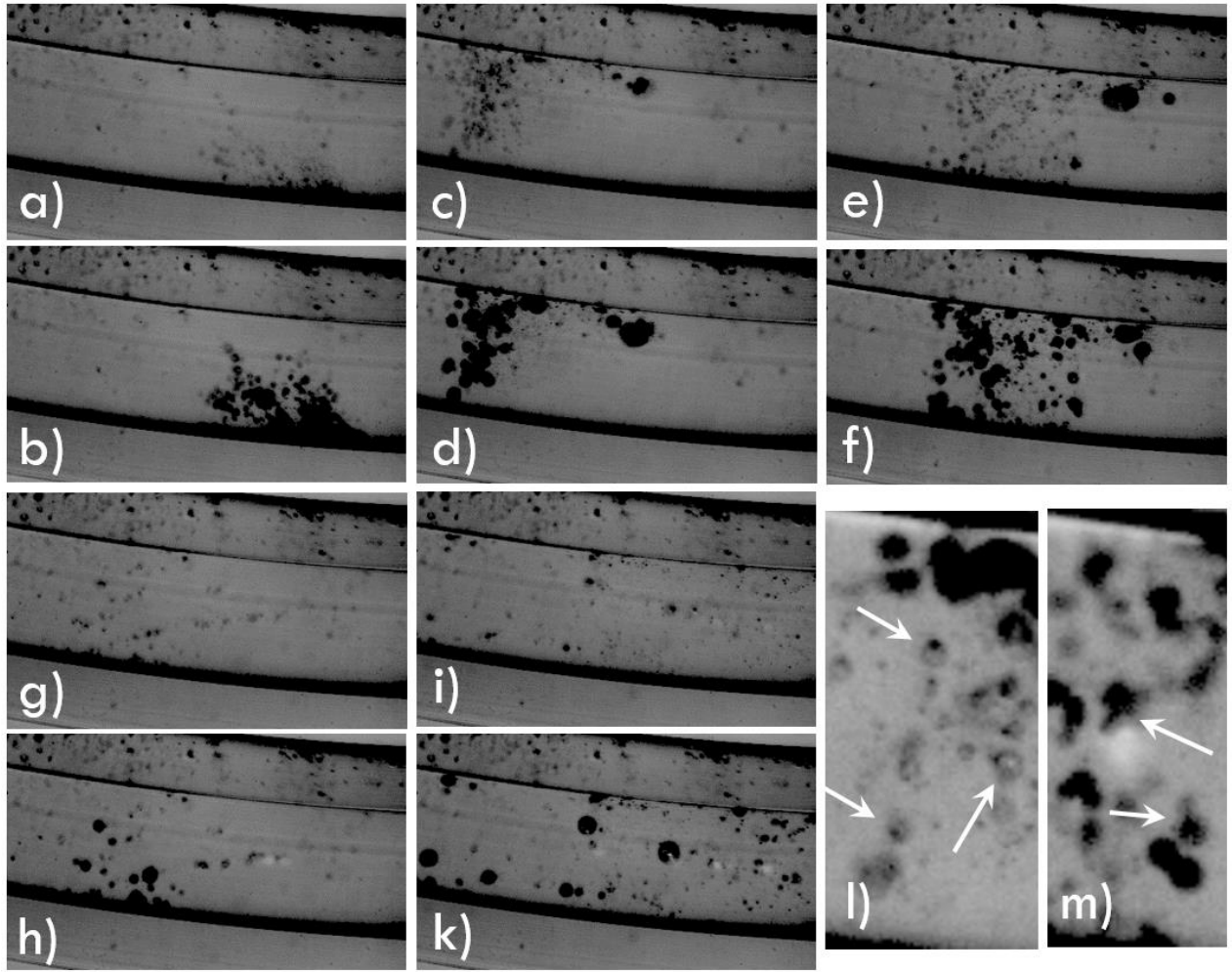

*Figure 4:* *Different cavitation bubble structures in the tube, all shown in a nearly collapsed phase and the subsequent phase near maximum expansion: Wall attached cluster (a, b); narrow plug (c, d); wider plug (e ,f); streamer (g, h); disperse (i, k). Displacement and jetting of collapsing bubbles are marked in frames l) and m). Recording with 20000 fps, exposure time 50 μs, frame heights ca. 1.5 mm in a) to k) and ca. 0.8 mm in l) and m). See also Movie1.avi in the supplementary material.*

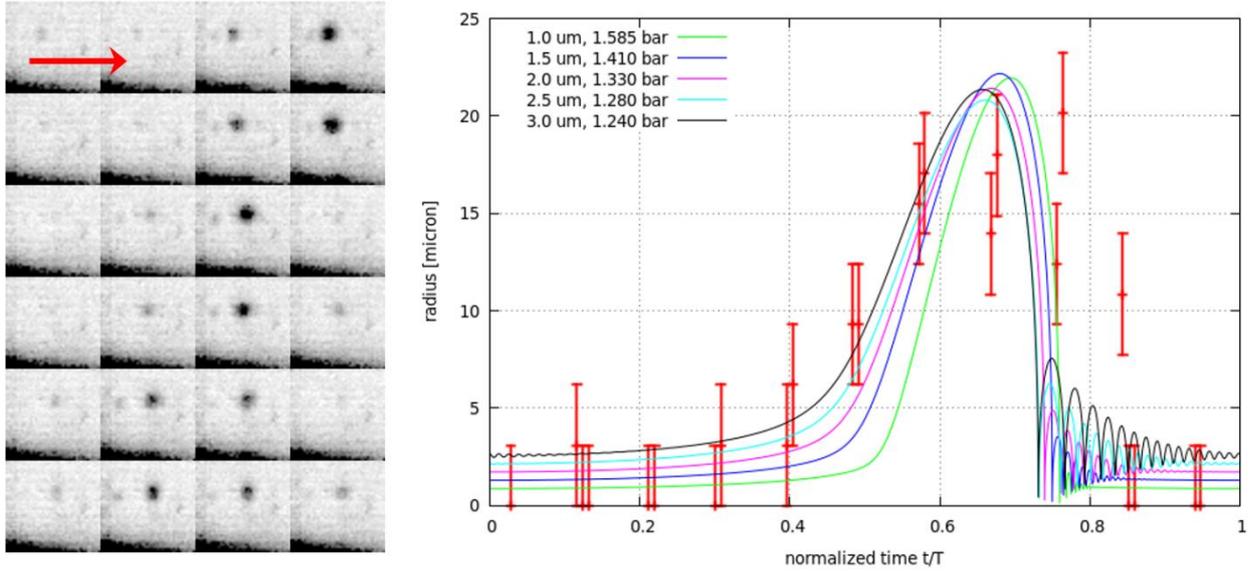

*Figure 5: A single bubble at the edge of a cluster inside the tube. Left: Image series showing the nearly stationary dynamics over 24 frames, corresponding to 6 driving periods (sequence from top left in direction of arrow, row by row; recording at 100 kfps, 10 µs exposure time, frame width 192 µm). Right: Reconstructed radius-time dynamics (experimental radius data with error bars, back-folded onto one driving period). Various combinations of equilibrium radius and driving pressure amplitude are indicated by color.*

**4.3 Bubble dynamics reconstruction**

Since the single-bubble model that is employed for the radius-time reconstruction is based on spherical and stationary oscillation, one should apply it to a bubble being more or less isolated for a few cycles. However, such bubbles are scarce within the structures, and not many test bubbles could be identified. Here we show a representative bubble at the border of a slim "plug", recorded at 100 kfps. Figure 5 shows on the left the section of the recording used, and the right plot gives the result of five fits for different equilibrium bubble sizes from $R_0 = 1.0$ µm to $R_0 = 3.0$ µm. This interval has been chosen since the bubble radius before large expansion corresponds roughly to the equilibrium radius, and during this state the bubble diameter is essentially at or below the spatial resolution limit of 6 µm. The pressure amplitudes leading to reasonable agreement of the numerical radius-time curves with the data range between 158.5 kPa for R=1.0 µm to 124.0 kPa for $R_0$=3.0 µm. From these parameters, the adiabatic compression temperatures can be derived from the numerically obtained radial compression ratio $\alpha = (R_0/R_{min})$ via $T = T_0 \alpha^{3(\gamma-1)}$. We obtain for the five fits values of alpha between about 7 ($R_0$=3.0 µm) and 8 ($R_0$=1.0 µm), which translates into peak temperatures between 3000 K ($R_0$=3.0 µm) and 3500 K ($R_0$=1.0 µm) in the collapse. As mentioned, these adiabatically heated bubble peak temperatures represent rather upper bounds, but the values should well be sufficient to dissociate trapped water vapor molecules to a substantial part into H and OH radicals [13] [43] [44]. Other bubbles in the structures show similar maximum and minimum radii as the particular bubble fitted here, which is why we conclude that luminol emission is consistent with the observed bubble dynamics in the clusters. Thus the localized bubble structures can unambiguously be identified as the sources of OH radicals and blue SCL light.

**4.4 Bubble nucleation**

The origin of cavitation bubble structures in the PFA tube is apparently based on nucleation events that occur at the free interface between liquid and gas slugs, since in the absence of the gas slugs, no cavitation is detected. Recordings in a setup virtually identical to Figure 1, but with horizontally aligned tubes in a holder frame, have captured individual bubble entrainments into the liquid from the gas. One such event is shown in Figure 6. The interface forms bulges and indentations, most likely connected to acoustically driven capillary waves [45]. Once seeded, the entering single bubble travels away from the interface and develops into a cluster by splitting and thus multiplying the bubble number. Calculation of the wavelength $\lambda_c \approx \sqrt[3]{\frac{2\pi\sigma}{\rho f_c^2}}$ [46] with the parameters for water and a Faraday capillary wave frequency of half the driving frequency, $f_c = f/2$, leads to $\lambda_c \approx 135$ µm. The observed bulge width in the center of the interface shown in Figure 6 amounts to about 85 µm, which is somewhat larger than the expected $\lambda_c/2 \approx 67.5$ µm. Probably, an influence of the spherical boundary conditions for the free surface inside the tube should be taken into account, leading to a surface oscillation mode of a wavelength different to the case of an infinite interface. The buildup of the bulge quite centrally on the axis of the tube gives further support for a symmetric mode oscillation of the interface here.

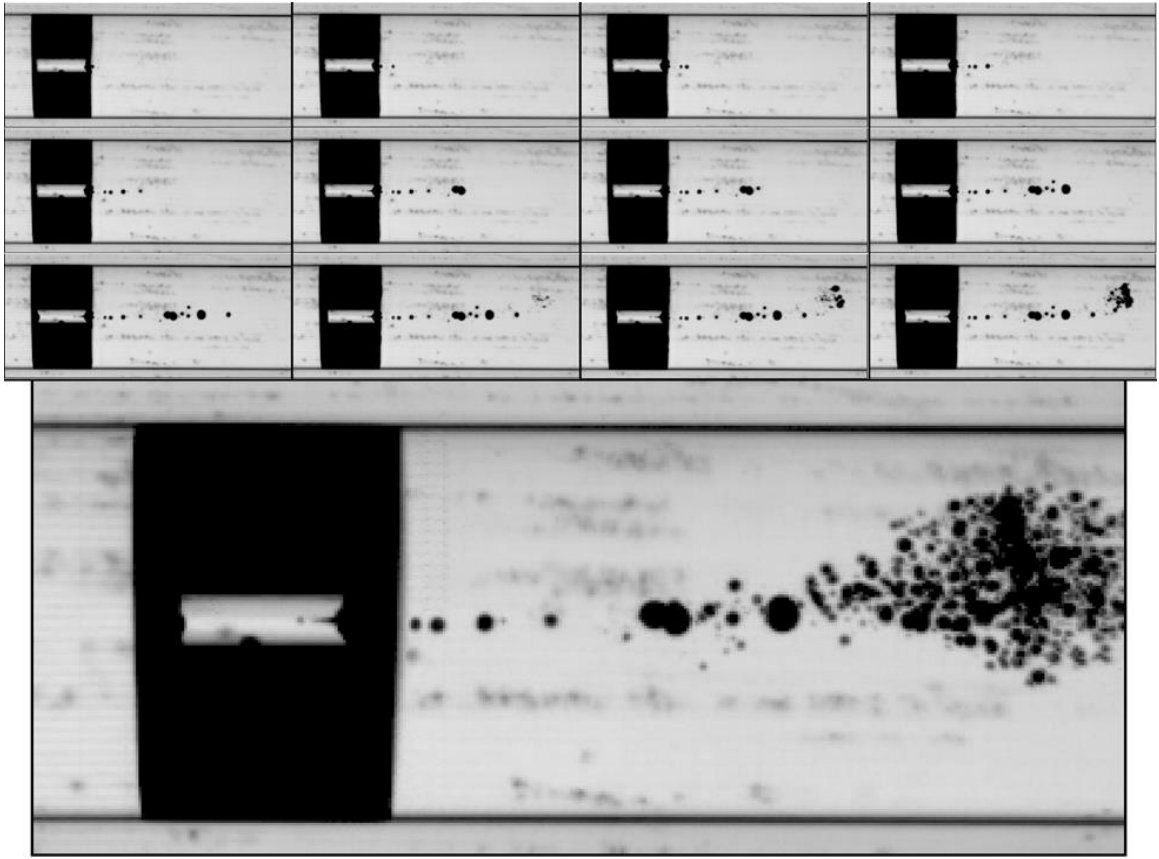

*Figure 6:* *Bubble nucleation at the free interface between gas and air slug: The upper picture series shows in form of cumulative images the path of the entering bubble (all former bubble positions stay visible in subsequent frames). The lower image shows a later stage when the bubble has transformed into a cluster (recording at 27500 fps, exposure time 1 µs, frame height 1 mm). See also Movie2.avi in the supplementary material.*

On the other side of the free interface, droplets can be ejected into the gas volume by essentially the same capillary wave dynamics. In Figure 7 such a case is presented were a central liquid jet is produced that disintegrates into drops. The first drop has a radius of about 11 µm (resulting in a volume of 5.6 pl). Its velocity reaches about 9 m/s, and the subsequent train of droplets flies with roughly 3 m/s into the gas slug. Ejected drops can hit the tube wall or the opposite interface of the next liquid slug. The drop ejection by capillary waves observed here appears similar to ultrasonic atomization at open free surfaces [47]. The atomization at inner free surfaces has also been observed to be responsible for the wetting of gas filled holes under ultrasound irradiation [48]. In our experiment, it seems that the slugs can change their length on a longer time scale due to the mass transfer by ejected drops and entering bubbles. From the image series in Figure 7 it appears that in this case the droplet ejection is also accompanied by a bubble creation. This is, however, not always happening. Still, drop ejection and bubble nucleation could be connected in some cases.

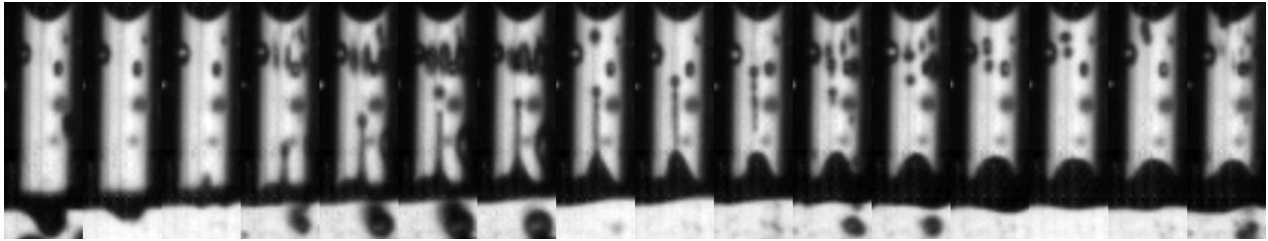

*Figure 7: Droplet ejection at the interface from water (lower part of image) into air (upper part). Recording at 150000 fps, exposure 2 µs, frame width approx. 200 µm, images turned 90° as compared to Figure 6. The bulge develops a thin jet that separates and disintegrates into drops. The dark structure below the interface after the third frame is a freshly created bubble that afterwards undergoes volume oscillations in the ultrasonic field. See also Movie3.avi in the supplementary material.*

## 5. Conclusion

We have investigated cavitation in 1/32" inner diameter PFA flow tubes submerged in an ultrasonic bath, running at 27200 Hz. Nucleation can only be observed under liquid/gas slug flow conditions where a free interface is present, which is in accordance to previous observation in directly irradiated microchannels [17] [45]. We have imaged nucleation events taking place by single bubble entrainment, induced by acoustically driven interface deformations, probably capillary waves. As well, droplets can be ejected into the gas phase via disintegration of liquid jets, apparently as well triggered by capillary waves. Bubble entrainment and drop ejection can happen simultaneously, as shown in one such event. Once nucleated, single entrained bubbles can develop into a larger bubble cluster. Generally, cavitation bubbles within the tube frequently form localized structures like clusters or plugs, i.e., confined small sections of the tube with cavitation activity. From individual bubble dynamics in a bubble structure, we estimate via numerical fitting the bubble equilibrium radius and the peak temperature. Obtained values in the range from 3000 K to 3500 K should be sufficient for a significant amount of hydrolysis and OH radical production of the water vapor trapped during collapse [49]. This is consistent with observations of SCL from luminol in the submerged sonicated tubes, indicating the presence of OH radicals.

In essence, we have confirmed that cavitation in flow tubes submerged in an ultrasonic bath can serve as a simple sonochemical flow reactor if bubble nucleation is facilitated. Apart from free interfaces, also other types of inhomogeneities might potentially be considered for bubble seeding [22]. The sonication of the tube via a coupling liquid might limit the reachable pressure amplitudes, in particular if cavitation and shielding in the coupling liquid occur. Furthermore, standing wave structures in the bath could inhibit cavitation in the full tube volume and thus shorten effective residence times. However, the presence of gas slugs in the tube might disturb the sound field and alleviate this effect. Future studies will focus on a better control of tube positions and field distribution, on more details of multi-bubble dynamics in the confined clusters, and on different coupling liquids.

**Acknowledgement**

The authors would like to thank the mechanical and electrical workshops at Drittes Physikalisches Institut for their support. The research leading to these results has received funding from the European Community's Horizon 2020 Programme [(H2020/2016 – 2020) under Grant Agreement no. 721290 (MSCA-ETN COSMIC)]. This publication reflects only the author's view, exempting the Community from any liability. Project website: https://cosmic-etn.eu/.